\documentclass[conference]{IEEEtran}
\IEEEoverridecommandlockouts

\usepackage{cite}
\usepackage{amsmath,amssymb,amsfonts}
\usepackage{algorithmic}
\usepackage{graphicx}
\usepackage{textcomp}
\usepackage{xcolor}
\usepackage{subcaption}
\usepackage{booktabs}
\usepackage{enumitem}
\usepackage{adjustbox}
\usepackage[hidelinks]{hyperref}
\usepackage{balance}
\usepackage{tikz}

\newcommand\copyrighttext{%
  \footnotesize \textcopyright \the\year{} IEEE. Personal use of this material is permitted. Permission from IEEE must be obtained for all other uses, including reprinting/republishing this material for advertising or promotional purposes, collecting new collected works for resale or redistribution to servers or lists, or reuse of any copyrighted component of this work in other works.}

\newcommand\copyrightnotice{%
\begin{tikzpicture}[remember picture,overlay]
\node[anchor=south,yshift=10pt] at (current page.south) {\fbox{\parbox{\dimexpr0.75\textwidth-\fboxsep-\fboxrule\relax}{\copyrighttext}}};
\end{tikzpicture}%
}
\def\BibTeX{{\rm B\kern-.05em{\sc i\kern-.025em b}\kern-.08em
    T\kern-.1667em\lower.7ex\hbox{E}\kern-.125emX}}
\begin{document}

\title{Enhancing Car-Following Models with Bike Dynamics for Improved Traffic Simulation}

\author{
\IEEEauthorblockN{Nico Ostendorf\IEEEauthorrefmark{1}\IEEEauthorrefmark{2},
Keno Garlichs\IEEEauthorrefmark{1},
Lars C. Wolf\IEEEauthorrefmark{2}
}

\IEEEauthorblockA{
\IEEEauthorrefmark{1} Corporate Research, Robert Bosch GmbH, Hildesheim, Germany \\
\IEEEauthorrefmark{2} Institute of Operating Systems and Computer Networks, Technische Universität Braunschweig, Braunschweig, Germany\\
\texttt{\{nico.ostendorf,keno.garlichs\}@de.bosch.com}, \texttt{wolf@ibr.cs.tu-bs.de}
}
}
\maketitle
\copyrightnotice

\begin{abstract}
Road traffic simulations are crucial for establishing safe and efficient traffic environments. 
They are used to test various road applications before real-world implementation. 
SUMO (Simulation of Urban MObility) is a well-known simulator for road networks and intermodal traffic, often used in conjunction with other tools to test various types of applications.
Realistic simulations require accurate movement models for different road users, such as cars, bicycles, and buses. 
While realistic models are already implemented for most vehicle types, bicycles, which are essential for achieving safe and efficient traffic, can only be modeled as slow vehicles or fast pedestrians at present. 
This paper introduces the Realistic Bicycle Dynamics Model (RBDM), the first dedicated bicycle model for SUMO, addressing this significant gap.
Leveraging real-world bicycle data from the SimRa dataset, the RBDM implements realistic speed, acceleration, and deceleration behaviors of bicycles in urban scenarios.
The evaluation is conducted using the Monaco SUMO traffic scenario (MoST) and a newly generated Berlin scenario in SUMO. 
The RBDM significantly outperforms the existing slow-vehicle approximation in SUMO, aligning more closely with real-world data across distribution, median, and interquartile range metrics.
These results underscore the necessity of a realistic bicycle movement model for accurate simulations, given the significant differences in the movement profiles of bicycles, cars, and pedestrians.

Furthermore, the model is tested for its ability to generalize to disparate scenarios and urban topologies, which is dependent on the manner and geographical region in which the SimRa data were gathered. 
In addition, recommendations are provided for how it could be adapted for use in different city topologies.
The enhanced realism of the RBDM is essential for accurately simulating and evaluating applications such as Vehicle-to-X communication and urban planning systems that depend on precise bicycle movement representation.
\end{abstract}


\section{Introduction}

Road traffic simulations are becoming increasingly important in the design of a safe and efficient traffic environment.
These simulations are essential for testing various applications and designs before implementing them in the real world. 
For example, they are used to study Vehicle-to-X (V2X) communication systems, and also in the design of road infrastructure, such as traffic lights and new roads.
Many of these tests and studies rely on the open-source simulator SUMO (Simulation of Urban MObility), which is capable of simulating road networks and various types of traffic participants, such as cars, buses, trains, bicycles, and pedestrians, each with their own dynamic properties \cite{Lopez2018Microscpic}. 
However, the accuracy of the dynamic properties for bicyclists is not as high as that of other road participants, which is particularly important given the increasing importance of bicycles in infrastructure planning and V2X communication \cite{Lobo2022Enhancing}.
Currently, cyclists are represented by a model that treats them as either slow cars or fast pedestrians.
Therefore, it is crucial to simulate bicycle dynamics in a more realistic manner, especially in simulations where authentic bicycle movement is essential, such as in V2X simulations. 
The accuracy of V2X simulations is highly dependent on a realistic bicycle model, as the message generation process for bicycle V2X messages is based on the movement of the bicycle. 
This process depends on factors such as current speed, acceleration, deceleration, heading, and the movement of neighboring vehicles.

This paper presents the Realistic Bicycle Dynamics Model (RBDM) as the first dedicated bicycle Car-Following-Model (CF-Model) in SUMO.
CF-Models focus on the longitudinal behavior of vehicles. 
The current implementation of bicycles as slow cars or fast pedestrians can be attributed to the limited availability of data for cyclists compared to data for cars or other vehicles. 
However, projects like SimRa \cite{Karakaya2020SimRa} are beginning to address this issue by providing an increasing amount of data on cyclists. 
This includes crucial parameters such as speed, position, accuracy, accelerations, and more, collected from various rides in Berlin.
To create a CF-Model for bicycles, our study used these real-world bicycle data from the SimRa data sets. 
Our contribution therefore involves filtering of the data to ensure high quality, analyzing it for speed, acceleration, and deceleration behavior, and implementing a CF-Model for bicycles in SUMO based on the results. 
We used different simulation scenarios to compare our bicycle model with the current implementation of bicycles in SUMO, in relation to the SimRa real-world data.
The first scenario was a newly created Berlin scenario, enabling us to evaluate the implemented RBDM in a setting comparable to the SimRa data. 
The second scenario was the well-evaluated Monaco SUMO traffic scenario (MoST) \cite{Codeca2017Most}, chosen to evaluate the model in a distinctly different environment. 
This allowed us to assess whether the model's behavior of bicycles would be different in cities with different topologies. 

The model is available to the research community at \url{https://github.com/boschresearch/RealisticBicycleDynamicsModel} under opensource licenses.

The remainder of this paper is structured as follows:
Section \ref{sec::relatedWork} provides an overview of related work. 
The fundamentals of SUMO and SimRa are explained in Section \ref{sec::Basics}. 
The evaluation of the SimRa data set is detailed in Section \ref{sec:SimRaBehavior}. 
The bike model is elaborated on in Section \ref{sec::bikeModel}.
Section \ref{sec::eval} includes the evaluation of the bike model. 
Finally, Section \ref{sec::Conc} presents the conclusion.

\section{Related Work}
\label{sec::relatedWork}
Karakaya et al. \cite{Karakaya2023Achieving} concentrate on the improvement of the existing bicycle modeling in SUMO with a large real-world data set of bicycle trips. 
For a realistic modelling of bicycles the bikes where divided in three different types of drivers: fast, medium and slow driver. 
Each group selects the existing SUMO parameters for maximum acceleration, maximum deceleration and maximum velocity based on a distribution function instead of a scalar value. 

An additional objective was to enhance the left-turning behavior of the bicycles in the simulation. 
To achieve this, a left-turn distribution was computed to determine the probability of cyclists choosing either the direct or indirect path when crossing the intersection.

The primary distinction of our research is the implementation of a new CF-Model for bicycles instead of using the existing model with different parameters. 
Therefore, we do not set maximum values for deceleration and acceleration, but select the current acceleration based on the distributions in the real world and situation. 
It is expected that choosing the actual deceleration and acceleration rather than the maximum value based on the distribution function will result in more realistic behavior.

Kaths and Grigoropoulos \cite{Kaths2016Modeling} concentrate on the implementation, evaluation and calibration of four different models to describe the acceleration and deceleration processes of bicyclists.
The four models presented are the constant model, the linear decreasing model, the two term sinusoidal model and the polynomial model.
To evaluate the different models, 1030 trajectories were used from four intersections in Munich. 
Based on the acceleration dynamics, the polynomial model was the most flexible and the model that estimated the acceleration profile best.
However, the maximum acceleration and deceleration were underestimated by this model. 
The two term sinusoidal model gave good estimates of the overall acceleration and deceleration profiles. 
And also for the maximum values of deceleration and acceleration.
The constant model was the best for both acceleration and deceleration. 
All three of these models could give good results based on the selected application.
The linear decreasing model gave the worst estimates of the acceleration profiles. 
Therefore, its use is not recommended compared to the other models. 

The main difference compared to our work is that we consider speed in addition to acceleration and deceleration. 
Furthermore, our model considers not only intersection areas, but entire bicycle routes in Berlin.
The data set we use is also much larger.

Heinovski et al. \cite{Heinovski2019Modeling} developed a virtual cycling environment to record and study cycling behavior. 
The environment consists of a real bicycle, a training stand and a virtual reality headset.  
The recorded data can be integrated directly into SUMO and is used to evaluate different warning systems. 

This paper also concentrates on realistic bike behavior in a simulation, but does not create a universal model for different types of cyclist. 
Therefore the system may not be suitable for some use cases due to scalability limitations. 
For example, conducting channel congestion studies requires a large number of bicycles with realistic behavior, making it impractical to scale such a simulator. 
Similarly, road infrastructure planning requires a large number of bicycles within a simulation to accurately represent real-world scenarios.
This scalability challenge represents a key difference in our research.

The papers \cite{Treiber2000Congested, Pourabdollah2017Calibration, Laquai2013Multi, Krauss1997Metastable, Song2014Research} primarily concentrate on the implementation and enhancement of models for realistic and safe movement in simulation, with a specific focus on vehicles such as cars. 
Some of these works also focus on very specific topics, such as a model for cooperative advanced driver assistance systems \cite{Laquai2013Multi}. 
However, none of these papers address the development of a movement model for bicyclists, which sets our paper apart.

\section{Background}
\label{sec::Basics}
\subsection{SUMO}
\label{sec::Sumo}

SUMO is an open-source microscopic and continuous multimodal traffic simulation platform. 
The term ``microscopic'' refers to the individual dynamics of each vehicle within the simulated scenario, allowing for highly detailed and realistic simulations.
SUMO scenarios use network data such as roads, cycle paths and buildings, as well as additional infrastructure like traffic lights, and a set of vehicles, e.g., cars and bicycles. 
For evaluations, SUMO provides output files containing extensive information like vehicle trajectories and traffic data \cite{Lopez2018Microscpic}.

In SUMO, each road participant is assigned a specific vehicle type (vType), which possess various attributes such as maximum speed, acceleration, and deceleration. 
Users have the flexibility to create custom vTypes or utilize predefined ones. 
A vType also includes a CF-Model and a Lane-Change Model (LC-Model). 
The CF-Model governs the longitudinal kinematic behavior, while the LC-Model controls the lateral kinematic behavior. 
While these models are well defined for cars and other motorized vehicles, there is currently no dedicated model for bicycles within SUMO. 
Although it is possible to define a vType with the vehicle class bicycle, the behavior of such bicycles is limited.
They will always behave as a fast pedestrian or a slow car \cite{Sumo2024Website}.

\subsection{SimRa}
\label{sec::SimRa}
The SimRa project is an open source research initiative designed to improve cyclist safety by collecting and analyzing data related to bicycle accidents and near misses.
One of the key components of the SimRa project is the SimRa app.
The app captures various data points, including location, time, motion sensor readings, road conditions, weather, near miss incidents, and traffic volume.
This information is then anonymized and aggregated to form a comprehensive data set for analysis and research purposes \cite{Karakaya2020SimRa}.

The data set comprises over 90,000 trips across Germany, with more than 35,000 publicly available trips specifically in Berlin. 
Our paper focuses on the publicly available trips in Berlin \cite{Bermbach2019Simra,Bermbach2021Simra,Bermbach2022Simra,Bermbach2023Simra}, with a special focus on the positioning and motion sensor data.

\section{Cycling behavior in SimRa}
\label{sec:SimRaBehavior}
To develop a realistic bicycle model in SUMO, our initial step involves analyzing real bicycle data from the SimRa data set. 
We begin by filtering the data to ensure high data quality. 
Subsequently, we examined the speed behavior of the bicycles in the SimRa dataset, followed by an analysis of their acceleration and deceleration behavior. 
Based on the insights gained from this analysis, we proceed to implement these longitudinal behaviors in an LC-Model as our RBDM.

\subsection{Preprocessing the data}

The SimRa data sets are derived from smartphone data, and it is important to note that the position information recorded by smartphones may not be entirely accurate \cite{Zangenehnejad2021Gnss}. 
To ensure the reliability of the data, we implemented a filtering process.
All samples utilized for the evaluation meet the following criteria:
\begin{itemize}
    \item The recorded sample includes GNSS accuracy information.
    \item In $90\%$ of cases, the GNSS accuracy is below $20\,m$.
    \item The average GNSS accuracy is below $12\,m$.
    \item The speed is below $16.67\,m/s$ in $95\%$ of the cases.
    \item The average speed is below $10\,m/s$.
    \item Each sample includes at least five acceleration and deceleration phases.
    \item The acceleration is in $90\%$ of the cases below $3\,m/s^2$.
    \item The deceleration is in $90\%$ of the cases above $-3m/s^2$.
\end{itemize}

To ensure high data quality, it is crucial to include only samples with GNSS accuracy information. 
We utilized all samples with an accuracy below $20\,m$ in $90\%$. 
This represents a trade-off between accuracy and the quantity of usable samples. 
The same trade-off applies to the average accuracy of below $12\,m$. 
Furthermore, opting for more precise values in conjunction with other filtering mechanisms will lead to a low number of usable samples.

Speeds above $16.67\,m/s$ are highly unlikely for bicycles, particularly in an urban scenario. 
A high number of such speeds may indicate measurement errors or the use of the measurement app on a vehicle other than a bicycle. 
Therefore, we filter out those tours. 
The same applies to average speeds, as average speeds above $10\,m/s$ are unrealistic for tours in a densely populated urban scenario like Berlin, especially when considering stops, such as at traffic lights.

In order to obtain only samples that represent meaningful trips with accelerations and decelerations, we have chosen to filter out trips with less than five accelerations and decelerations.
A deceleration of more than $3m/s^2$ already indicates strong braking and may suggest dangerous situations. 
A high number of such decelerations could also indicate measurement errors for the current speed. 
The same criteria apply to acceleration, as an acceleration of more than $3,m/s^2$ is already very high for bicycles. 
A high percentage of such accelerations could also indicate measurement errors.

After applying the explained filtering process, $84\%$ ($34,273$) of all tours remain and are available for evaluation. 
This rigorous filtering ensures that the data set used for analysis meets specific quality standards.

\subsection{Speed distribution}
\label{sec::SimRaSpeedDist}
\begin{figure}[ht]
\includegraphics[width=\linewidth]{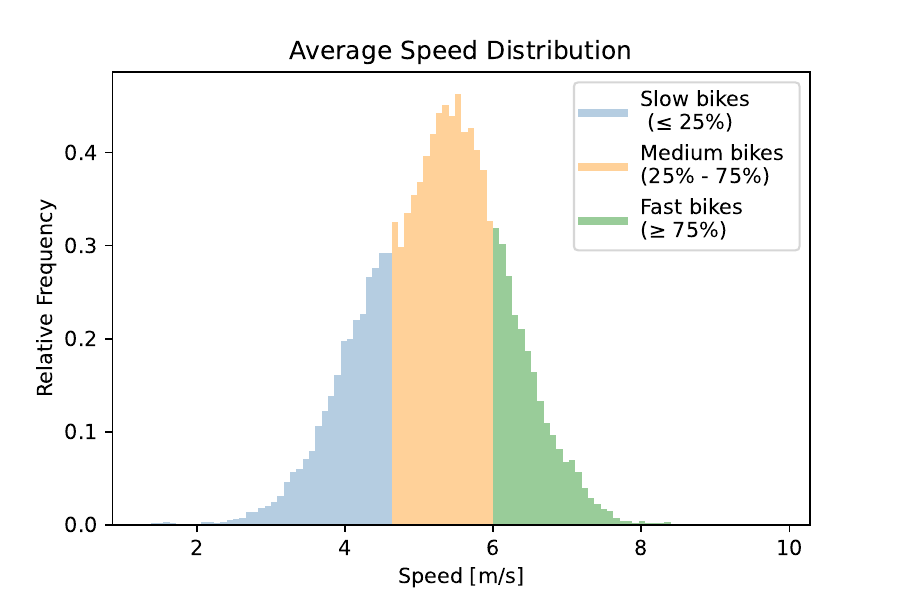}
\caption{Distribution of average speeds of bikes in the SimRa data set. Split into three groups.}
\label{fig:avgSpeedDist}
\end{figure}

In the initial analysis, the focus is on the average speed of the various tours, considering that the standstill time of each bicycle is significantly affected by current traffic and traffic lights. 
Therefore, the decision was made to calculate the average speed based solely on movement, selecting speeds exceeding $1\,m/s$ for a more accurate comparison with the later simulated scenario. 
This was because the data set lacks information about current traffic, making it impossible to replicate exact traffic conditions in our simulations.
The result is plotted in Figure \ref{fig:avgSpeedDist}. 

We assume that different groups of bicyclists will behave differently in the context of acceleration and deceleration. 
For this reason we decided to group the different drivers based on their speeds. 
The average speeds are divided into three groups. 
Slow drivers, representing the slowest $25\%$ of all tours, maintain an average speed between $1.3\,m/s$ and $4.6\,m/s$. 
Fast drivers, which make up the fastest $25\%$ of all tours, achieve an average speed between $5.9\,m/s$ and $9.8\,m/s$. 
The medium speed drivers fall in between, with an average speed ranging from $4.6\,m/s$ to $5.9\,m/s$. 
Overall, the mean average speed of bicycles in this data set is $5.3\,m/s$.
We have opted for a 25\%-50\%-25\% split because we believe that the majority of riders will be best represented by the medium bikes, with a smaller proportion of riders representing either the fast or the slow bikes.

\begin{figure}[ht]
\includegraphics[width=\linewidth]{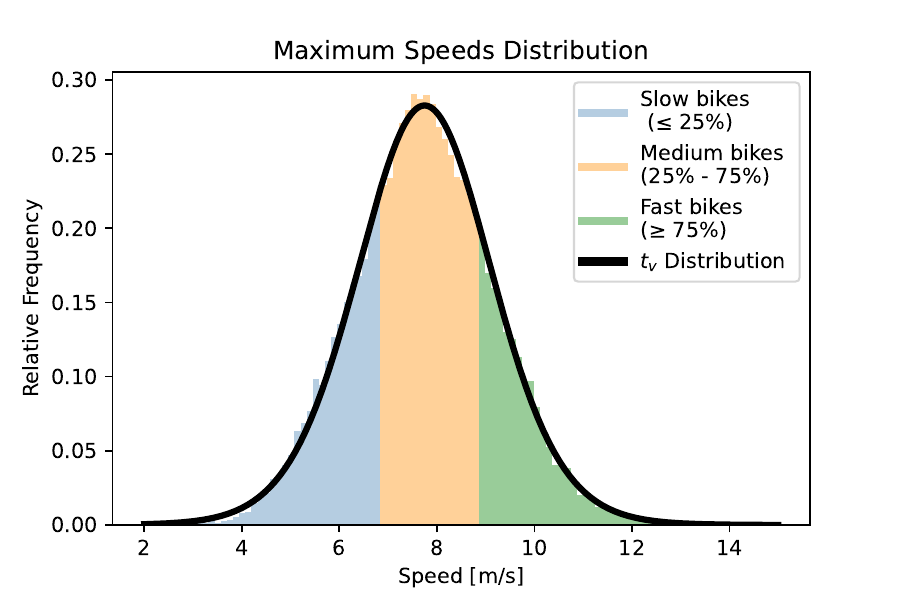}
\caption{Distribution of maximum reached speeds of bikes in the SimRa data set. Split into three groups.}
\label{fig:maxSpeedDist}
\end{figure}

Another crucial evaluation involves determining the maximum speeds achieved during each trip.
To ensure accuracy, we have selected the maximum speed that is sustained for a minimum of $10\,s$.
With this, we want to exclude values that are solely the result of measurement errors or similar causes.
This maximum speed is a crucial factor, particularly as every SUMO vType includes an attribute for the maximum speed of a vehicle.
The results are presented in Figure \ref{fig:maxSpeedDist} and are categorized similarly into three groups. 

Slow drivers typically reach a maximum speed ranging from $2.3\,m/s$ to $6.8\,m/s$, while fast drivers achieve speeds between $8.7\,m/s$ and $14.9\,m/s$.
Meanwhile, medium drivers achieve a maximum speed between $6.8\,m/s$ and $8.7\,m/s$. 
On average, across all trips, the maximum speed reached is $7.8\,m/s$. 

To implement the speed behavior in SUMO, we intend to use a probability density function (PDF) for the distribution of maximum speed values.
To determine the best fitting PDF, we conducted tests using various PDFs with different values and selected the PDF with the lowest sum squared error (SSE), which represents the summed difference between the observed and predicted values. 
As a result, the distribution of maximum speeds is represented by the Student's t distribution $t_v$ \cite{Student1908Probable}:
$$f(x,\nu) = \frac{\ \Gamma\left( \nu+1/ 2 \right)\ }{\ \sqrt{\pi\ \cdot \nu\ }\cdot \Gamma\left(\nu/2\right)} \left(\ 1 + x^2/\nu\ \right)^{-(\nu+1)/2}$$  
Where $x$ is a real scalar and $\Gamma$ is the gamma function. 
To fit $t_v$ to our distribution we need to shift and scale the function using the $loc$ and $scale$ parameters.
This leads to: $$\frac{f(x', \nu)}{scale} \text{ with } x' = \frac{(x - loc)}{scale}$$
The specific parameters are: $\nu = 17.287$, $loc = 7.748$ and $scale = 1.39$.

\subsection{Acceleration}
\begin{figure}[ht]
\includegraphics[width=\linewidth]{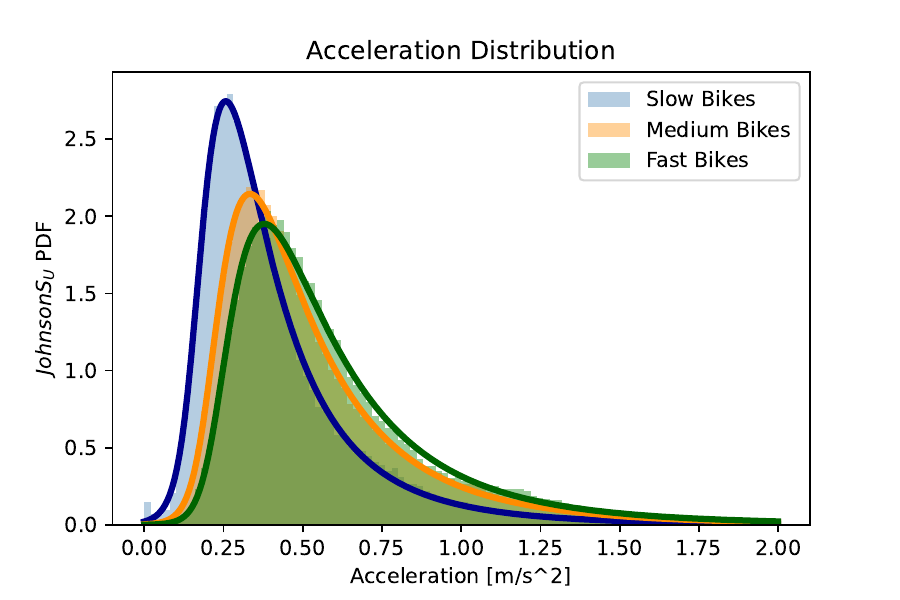}
\caption{Distribution of the accelerations of bikes in the SimRa data set. Split into three groups.}
\label{fig:AccelSimRa}
\end{figure}

Another crucial parameter for achieving realistic cyclist behavior is acceleration. 
The distribution of acceleration in the SimRa data set is illustrated in Figure \ref{fig:AccelSimRa}. 

As in the previous speed analyses, the distributions are segmented into three distinct groups. 
The groups are based on the maximum speed reached by the bicycle, as in Figure \ref{fig:maxSpeedDist}.
The distribution of acceleration varies noticeably across these groups, with the slow speed group showing lower acceleration compared to the medium speed group, and the medium speed group displaying less acceleration than the fast speed group. 
Specifically, the slow drivers have a median acceleration of $0.34\,m/s^2$, while the medium drivers have a median acceleration of $0.45\,m/s^2$, and the fast drivers have a median acceleration of $0.51\,m/s^2$.

In this case, we also use a PDF for the distribution of acceleration values to implement realistic acceleration behavior in SUMO. 
Once more, we selected the PDF with the lowest SSE.
Notably, the acceleration of all speed groups conforms to the Johnson's $S_U$-distribution \cite{Johnson1949Systems}, represented by the equation:
$$f(x,a,b) = \frac{b}{\sqrt{x^2+1}} \cdot \phi(a+b \cdot log(x+\sqrt{x^2+1}))$$
Where $x, a, b \in \mathbb{R}$ and $\phi$ is the probability density function of the normal distribution. 
To fit Johnson's $S_U$-distribution to our distribution, we need to shift and scale the function using the $loc$ and $scale$ parameters.
This leads to:
$$\frac{f(x',a,b)}{scale}\text{ with }x' = \frac{(x - loc)}{scale}$$
The specific values for $loc$, $scale$, $a$, and $b$ for the different groups are detailed in Table \ref{tab:accelPDF}.

\begin{table}[ht!]
\centering
\caption{Parameters for Johnson's $S_U$ distribution for the acceleration of the different speed groups.}
\begin{tabular}{ccccc}
\toprule
\textbf{Group} & \textbf{$a$} & \textbf{$b$} & \textbf{$loc$} & \textbf{$scale$} \\\midrule
\textbf{Slow}& $-1.676$ & $1.218$ & $0.151$ & $0.105$\\
\textbf{Medium}& $-2.019$ & $1.270$ & $0.179$ & $0.117$\\
\textbf{Fast}& $-2.189$ & $1.314$ & $0.196$ & $0.123$\\\bottomrule
\end{tabular}
\label{tab:accelPDF}
\end{table}

\subsection{Deceleration}

\begin{figure}[ht]
\includegraphics[width=\linewidth]{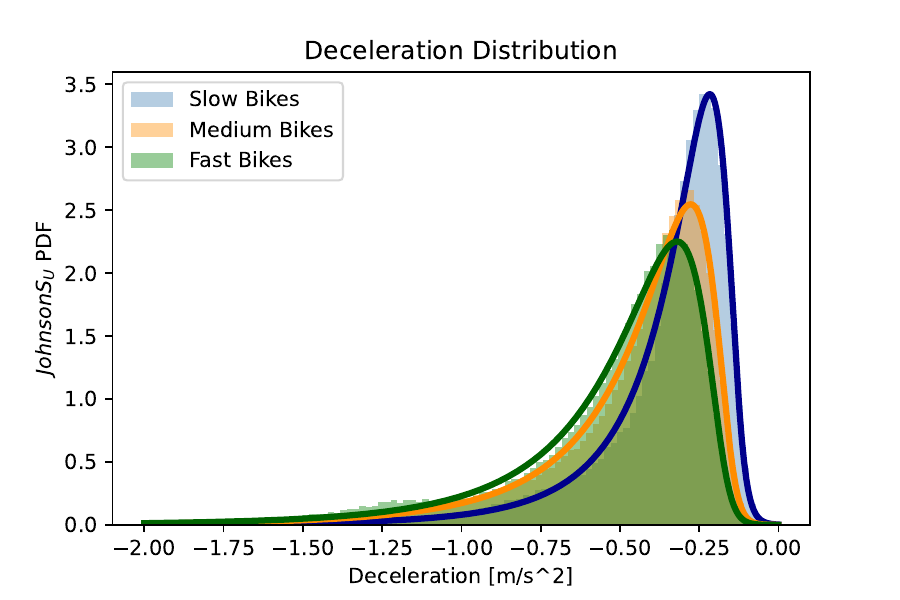}
\caption{Distribution of the deceleration of bikes in the SimRa data set. Split into three groups.}
\label{fig:DecelSimRa}
\end{figure}

The last crucial parameter to consider is deceleration. 
The distribution of the deceleration is plotted in Figure \ref{fig:DecelSimRa}.

Similar to the analysis of acceleration, the deceleration is segmented into the three speed groups. 
As expected, the deceleration behavior varies across the three groups, with the slow group exhibiting the lowest deceleration, the fast speed group displaying the highest deceleration, and the medium group falling in between. 
Specifically, the slow drivers have a median deceleration of $-0.29\,m/s^2$, while the medium drivers have a median deceleration of $-0.38\,m/s^2$, and the fast drivers have a median deceleration of $-0.44\,m/s^2$.

Again we use a PDF for the implementation of realistic deceleration behaviour in SUMO. 
Notably, the deceleration for all three groups conforms to the Johnson's $S_U$-distribution as the acceleration. 
The specific parameters for each group are detailed in Table \ref{tab:decelPDF}.

\begin{table}[ht!]
\centering
\caption{Parameters for Johnson's $S_U$ distribution for the deceleration of the different speed groups.}
\begin{tabular}{ccccc}
\toprule
\textbf{Group} & \textbf{$a$} & \textbf{$b$} & \textbf{$loc$} & \textbf{$scale$} \\\midrule
\textbf{Slow}& $2.053$ & $1.237$ & $-0.123$ & $0.068$\\
\textbf{Medium}& $2.428$ & $1.279$ & $-0.138$ & $0.076$\\
\textbf{Fast}& $2.538$ & $1.334$ & $-0.152$ & $0.088$\\\bottomrule
\end{tabular}
\label{tab:decelPDF}
\end{table}

\section{Implementing SUMO Bicycle Model}
\label{sec::bikeModel}
Our developed RBDM is based on SUMO's version 1.8, chosen for its compatibility with other simulators such as OMNeT++ \cite{Varga2010Omnet} and Veins \cite{Sommer2011Bidirectionally}. 
This combination of simulators enables the simulation of V2X applications based on movement in SUMO. 
The principles of our model are also extendable to the newest version of SUMO.

Our RBDM is divided into two main parts. 
The first part involves an adjustment of the base vehicle class in SUMO, while the second part consists of a newly created CF-Model using the new parameters and functions of the modified base vehicle class.

In the vehicle class, we implemented Johnson's $S_U$ density functions for decelerations and accelerations. 
The parameters used for the density function are determined by the maximum possible speed of the bicycle, with each bicycle having a different maximum speed value based on the Student's t distribution with the parameters of Section \ref{sec::SimRaSpeedDist}. 
Future acceleration and deceleration values are calculated in advance and saved in a queue for each dedicated bicycle.

In the CF-Model, the values from the queues are used to calculate the next speed and the time point when a bicycle needs to start braking based on the next deceleration values, distance, and speed difference to the object in front. 
The CF-Model always calculates the safe distance to the vehicle in front and the maximum safe speed with the help of the next decelerations, accelerations and the maximum speed value. 
If braking with the values of the calculated deceleration queue would lead to an accident due to unexpected behavior of the vehicle in front or a red traffic light, the bicycle will brake as hard as necessary until the queue of deceleration values aligns again. 
The CF-Model consistently utilizes the calculated queue of the bicycle for acceleration. 
However, in cases where the calculated acceleration would result in unsafe or to high speeds, a lower acceleration value is selected instead.

Our RBDM ensures realistic and safe movement of bicycles, allowing them to maintain a safe distance from other road participants. 
This safe distance can be adjusted using existing SUMO parameters such as minGap and tau.

\section{Evaluation}
\label{sec::eval}
\subsection{Setup}
\begin{figure}[ht]
\includegraphics[width=\linewidth]{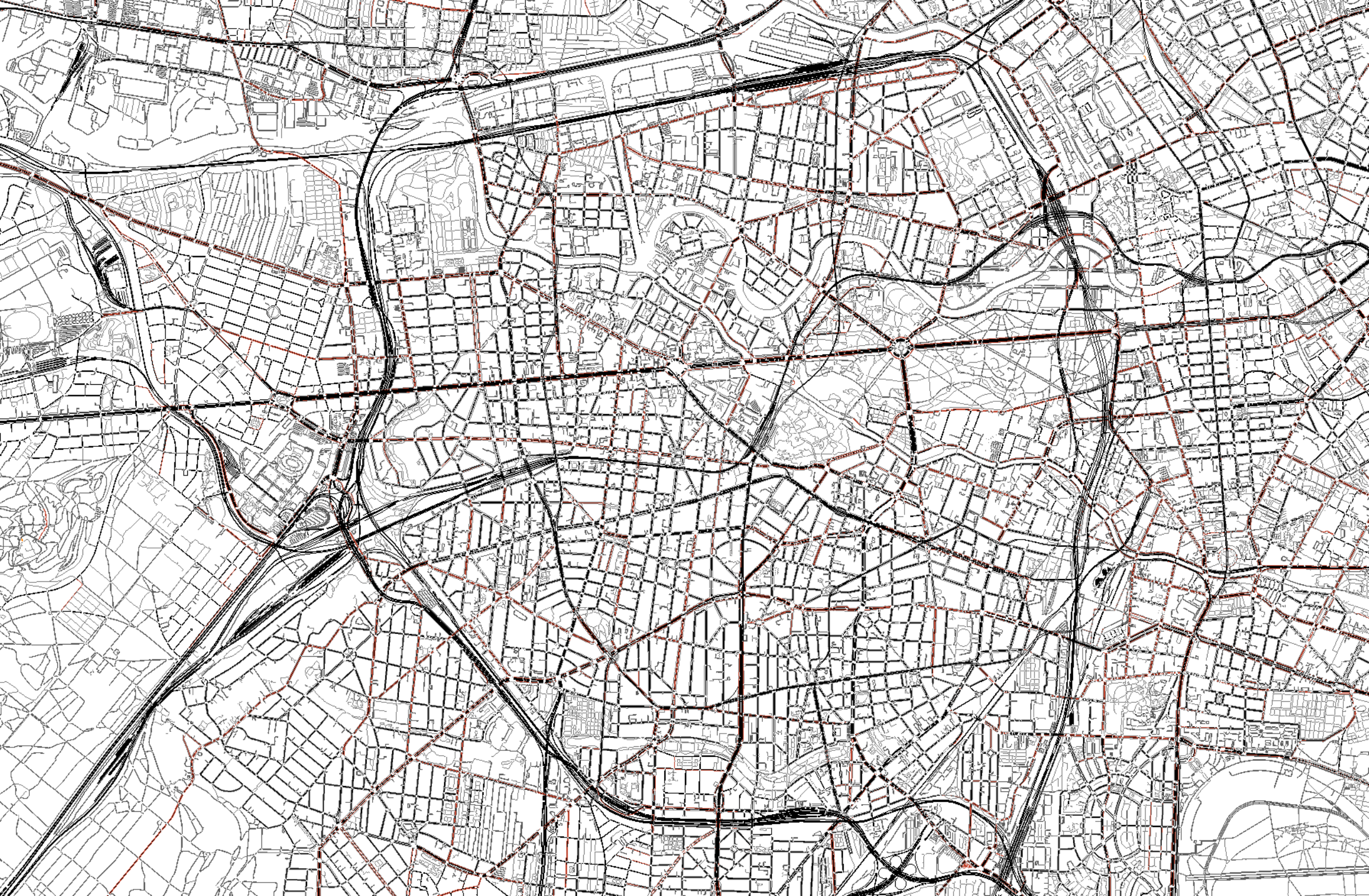}
\caption{SUMO Screenshot of Berlin OSM web wizard Scenario.}
\label{fig:berlin-Scenario}
\end{figure}

We are using two different scenarios to evaluate the developed bicycle model.
The first scenario involves a generated city-wide simulation of Berlin, inclusive of both bicycles and cars, created using SUMO's OSM web wizard which allows to create simulation scenarios from Open Street Maps.
The created scenario is displayed in Figure \ref{fig:berlin-Scenario}.
We created our own Berlin scenario, because the existing Berlin Sumo Traffic Scenario (BeST) \cite{Schrab2023LargeScale} does not include bicycles. 
Therefore, it is not possible to use this scenario for our evaluation.
Our Berlin scenario includes the sublane model of SUMO to reach realistic behavior of bicycles. 
The sublane model allows vehicles to overtake each other without leaving the main lane, if there is enough space. 
Vehicles will also wait next to each other instead of building a long queue at traffic lights. 
The scenario was specifically selected due to the recording of SimRa data in Berlin, ensuring a robust basis for comparison between the simulation results and the SimRa dataset.

\begin{figure}[ht]
\includegraphics[width=\linewidth]{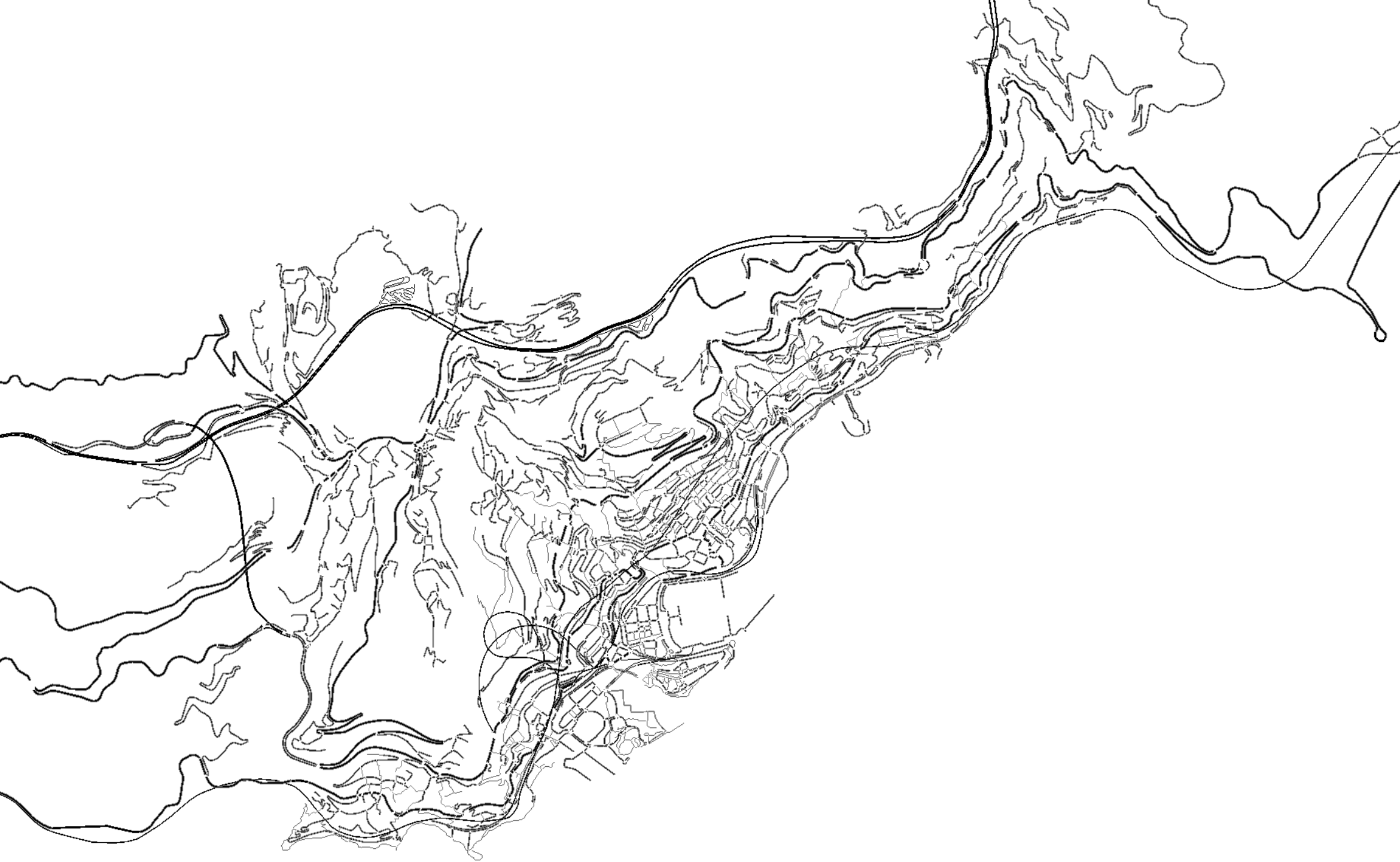}
\caption{SUMO Screenshot of Monaco Traffic Scenario.}
\label{fig:most-Scenario}
\end{figure}

The second scenario is the MoST\cite{Codeca2017Most} represented in Figure \ref{fig:most-Scenario}. 
This scenario includes more than 3000 bicycles and a realistic representation of the traffic of one day in Monaco. 
Given the differing topology between Monaco and Berlin, varying results are anticipated.

In both scenarios, we will evaluate the maximum speeds, the average speeds, the acceleration and the deceleration distribution for all bicycles. 
The results will be compared with those of the SimRa data set in Section \ref{sec:SimRaBehavior} and to the results of the simulation without the RBDM. 
The Berlin simulation without our RBDM implements bicycles as slow vehicles, which is the default setting when using the vType bicycle in SUMO.
The MoST scenario has three types of bicycles implemented by default: slow, average and fast bicycles. 
They are also implemented as slow vehicles, but each group of bicycles has different values for the maximum speed value.

\subsection{Berlin Scenario}

\begin{figure}[ht]
\begin{subfigure}[t]{0.23\textwidth}
    \centering
    \includegraphics[width=\linewidth]{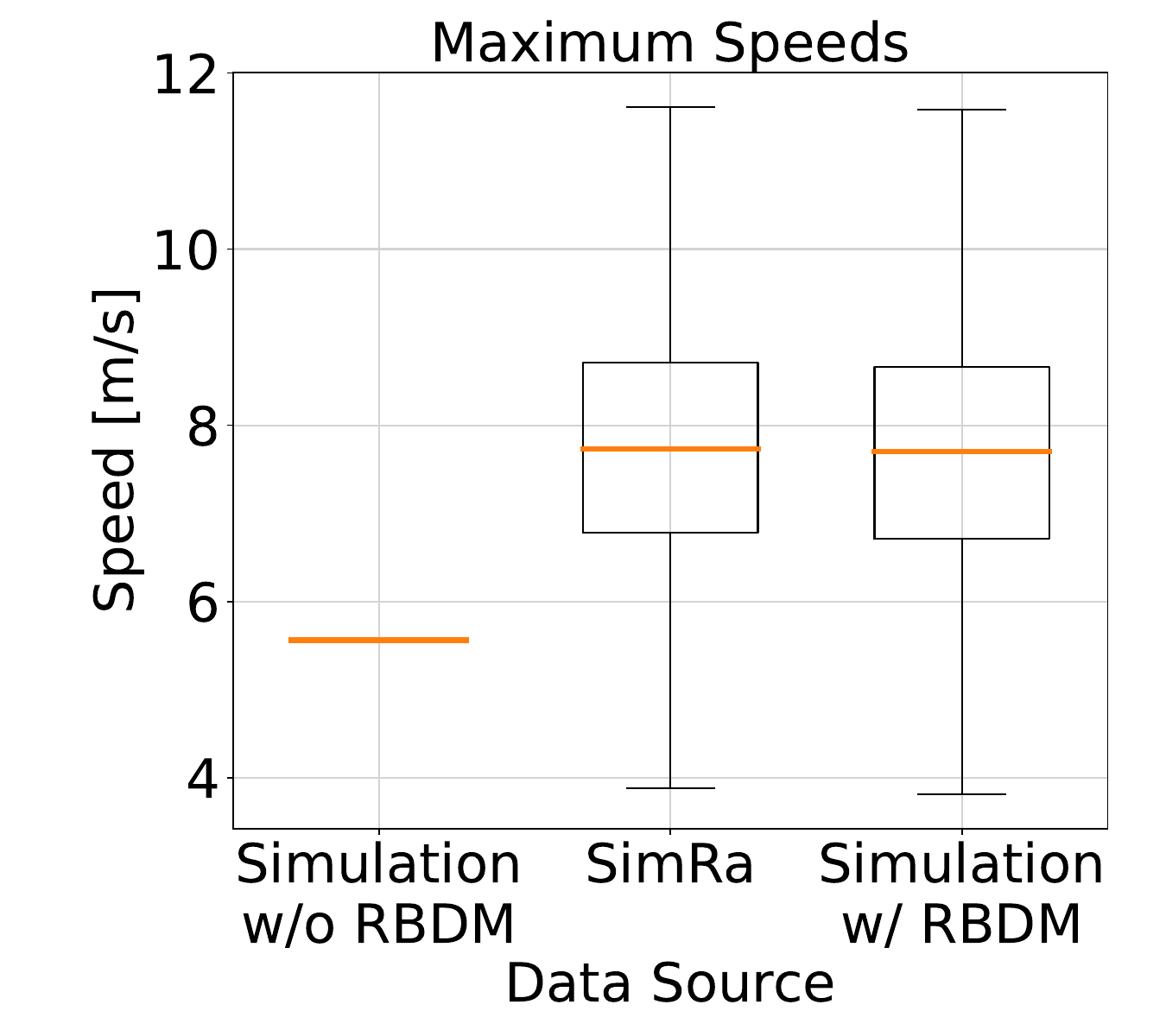}
    \caption{Maximum reached speed values of bicycles.}
    \label{fig:berlin-a}
\end{subfigure}
\hfill
\begin{subfigure}[t]{0.23\textwidth}
    \centering
    \includegraphics[width=\linewidth]{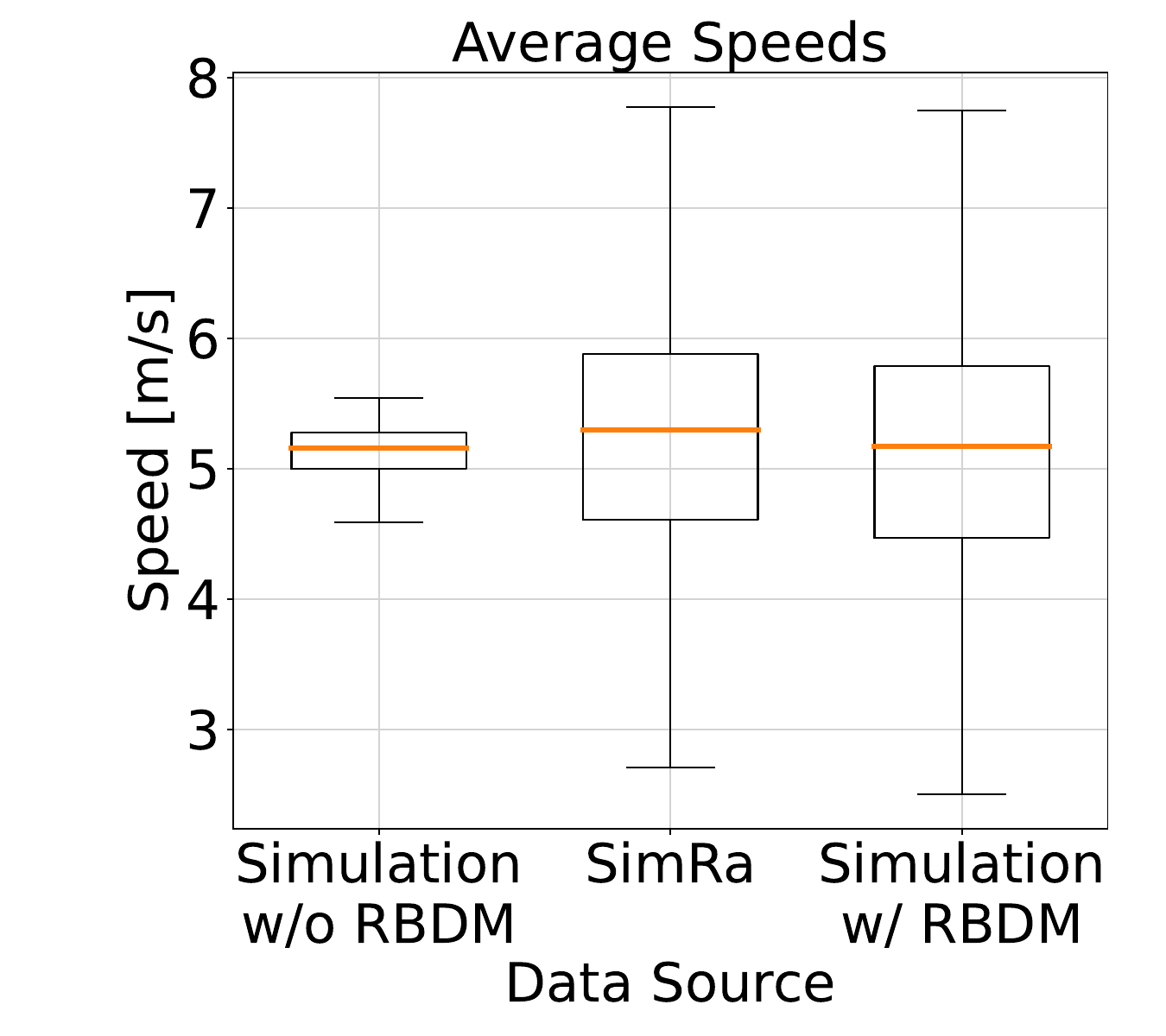}
    \caption{Average speed values of bicycles.}
    \label{fig:berlin-b}
\end{subfigure}
\hfill
\begin{subfigure}[t]{0.23\textwidth}
    \centering
    \includegraphics[width=\linewidth]{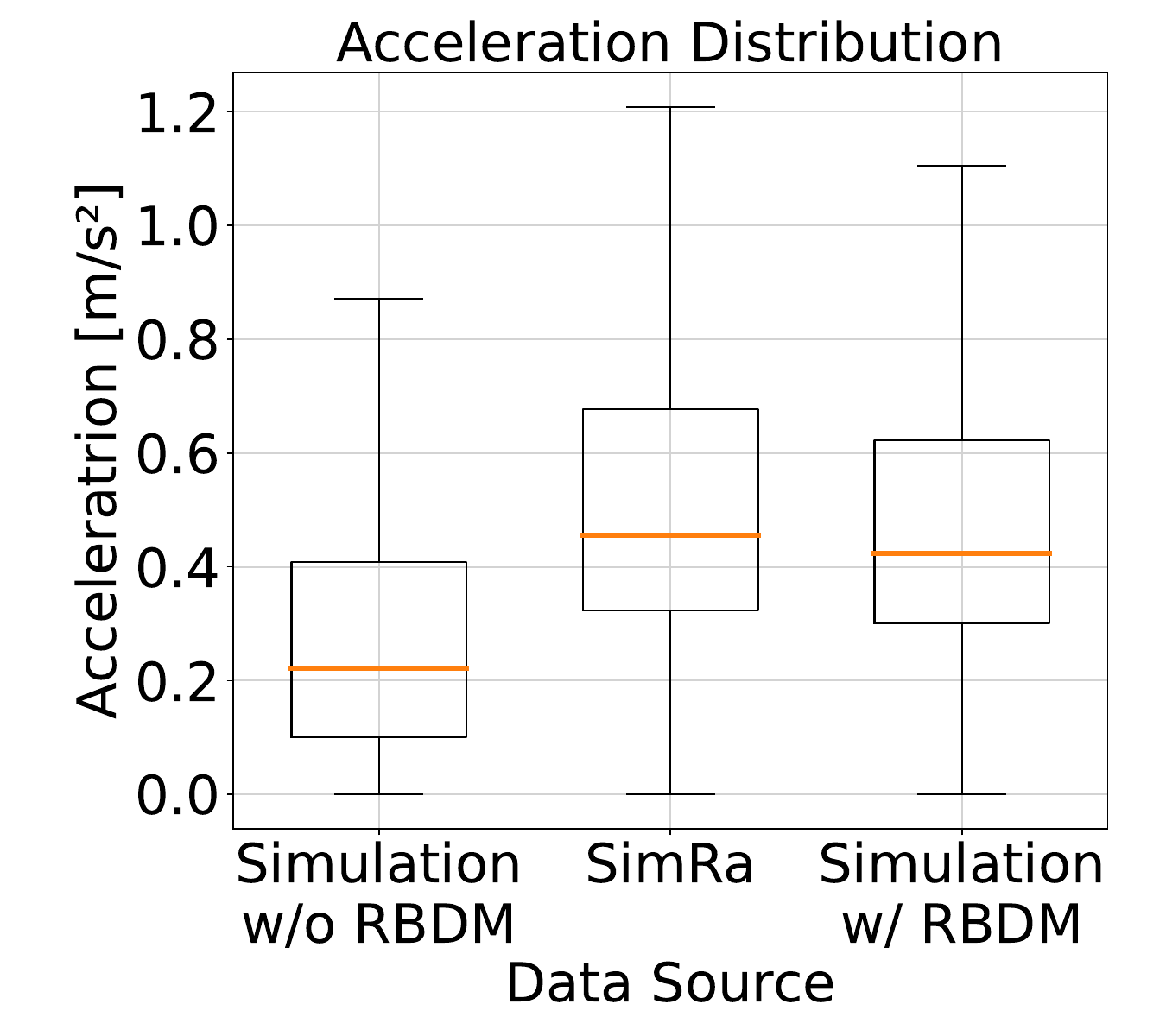}
    \caption{Distribution of acceleration of bicycles.}
    \label{fig:berlin-c}
\end{subfigure}
\hfill
\begin{subfigure}[t]{0.23\textwidth}
    \centering
    \includegraphics[width=\linewidth]{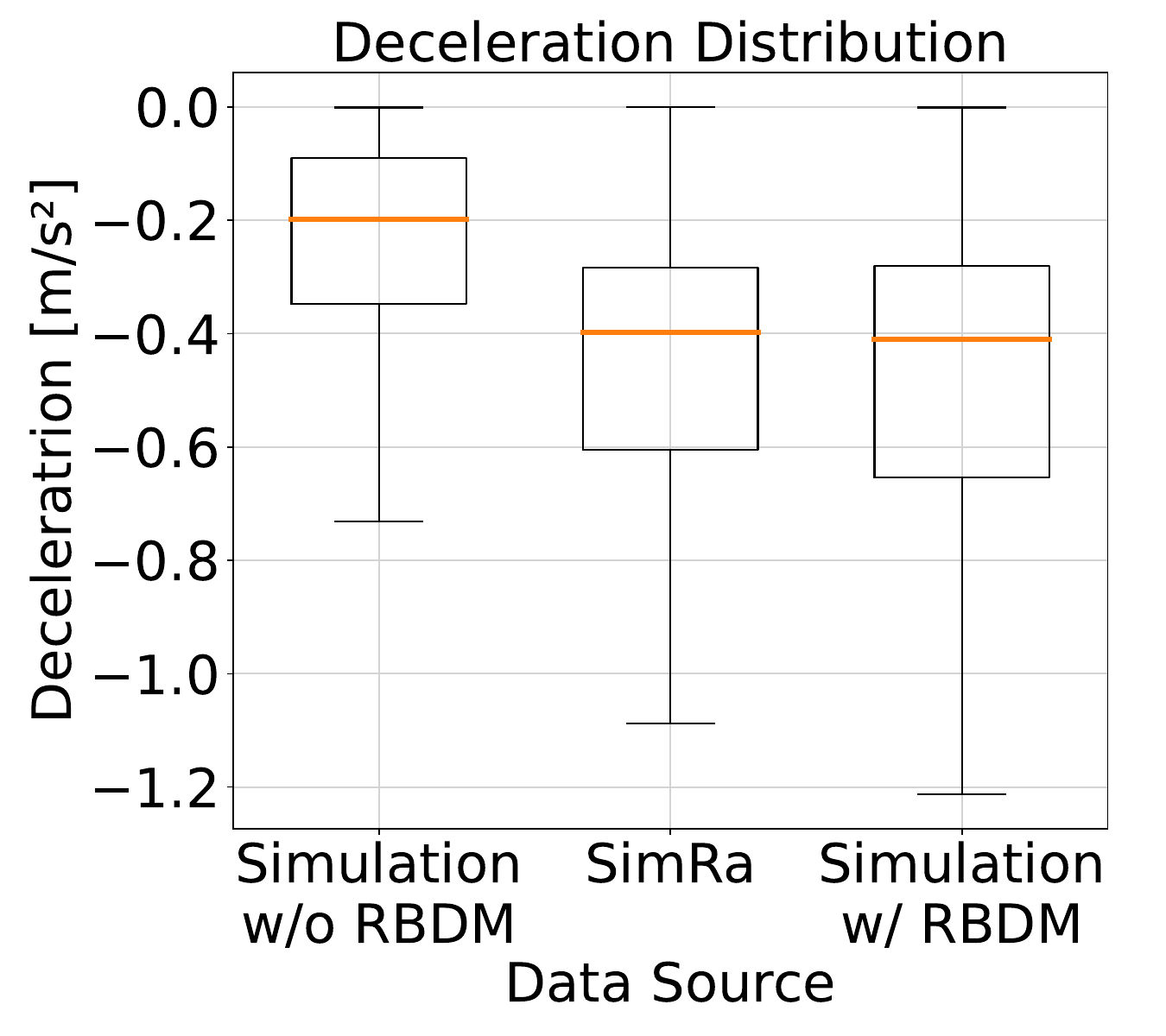}
    \caption{Distribution of deceleration of bicycles.}
    \label{fig:berlin-d}
\end{subfigure}
\caption{Results of Berlin scenario evaluation with and without our model compared to the SimRa data results.}
\label{fig:berlin}
\end{figure}

The results of the evaluation of the Berlin scenario are plotted in Figure \ref{fig:berlin} using boxplots. 
The median is shown by the orange line.
The upper and lower whisker of the boxplots are calculated as 1.5 times the interquartile range (IQR), with the maximum or minimum value being chosen if it is smaller than 1.5 times IQR.
We compared the simulated Berlin scenario with and without our RBDM to the results of the SimRa data.

Figure \ref{fig:berlin-a} represents the maximum speeds of all recorded bicycles for the different data sources. 
The simulation without our RBDM yielded a median value of $5.56\,m/s$, significantly lower than the SimRa data median value of $7.73\,m/s$. 
However, with RBDM, the median value improved to $7.7\,m/s$, representing a difference of less than $1\%$ from the SimRa data. 
Similarly, the maximum value also showed improvement with our model, with a difference of only $7\%$ compared to the SimRa data. 
The simulation with bicycles as slow vehicles leads to a $62\%$ lower maximum speed compared to the SimRa data.

Figure \ref{fig:berlin-b} represents the average speeds for the different data sources. 
While the median values were similar across all setups, IQR, 25th, and 75th percentile showed significant differences. 
For the SimRa data set the IQR is $1.27\,m/s$. 
The simulation without RBDM has a significantly lower IQR of $0.28\,m/s$, while the simulation with RBDM has a result similar to the SimRa data with $1.32\,m/s$. 
The simulation with RBDM demonstrated results closer to the SimRa data than the simulation without the model. 

In Figure \ref{fig:berlin-c} the acceleration distribution is presented, showing that the simulation with RBDM closely aligned with the SimRa data, while the simulation without our model differed significantly.
The median acceleration in the SimRa data is $0.46\,m/s^2$, in the simulation with RBDM $0.43\,m/s^2$ and in the simulation without RBDM $0.22\,m/s^2$. 

Lastly, Figure \ref{fig:berlin-d} displays the deceleration distribution, with the simulation using our model showing better alignment with the SimRa data compared to the simulation with bicycles as slow vehicles.
The median deceleration of our RBDM is $-0.41\,m/s^2$, near the median of $-0.4\,m/s^2$ in the SimRa data. In contrast, the simulation without our model yielded a median deceleration of $-0.2\,m/s^2$, which is $50\%$ lower than the SimRa data.

In summary, the simulation with our RBDM demonstrated better alignment with the SimRa data compared to the default setting for bicycles in SUMO. 
The speed variance and distribution, as well as the values for deceleration and acceleration, were notably improved with the use of our model. 
These results indicate the effectiveness of our model in enhancing the realism of bicycle behavior in simulations.

\subsection{Monaco SUMO Traffic Scenario}
\begin{figure}[ht]
\begin{subfigure}[t]{0.23\textwidth}
    \centering
    \includegraphics[width=\linewidth]{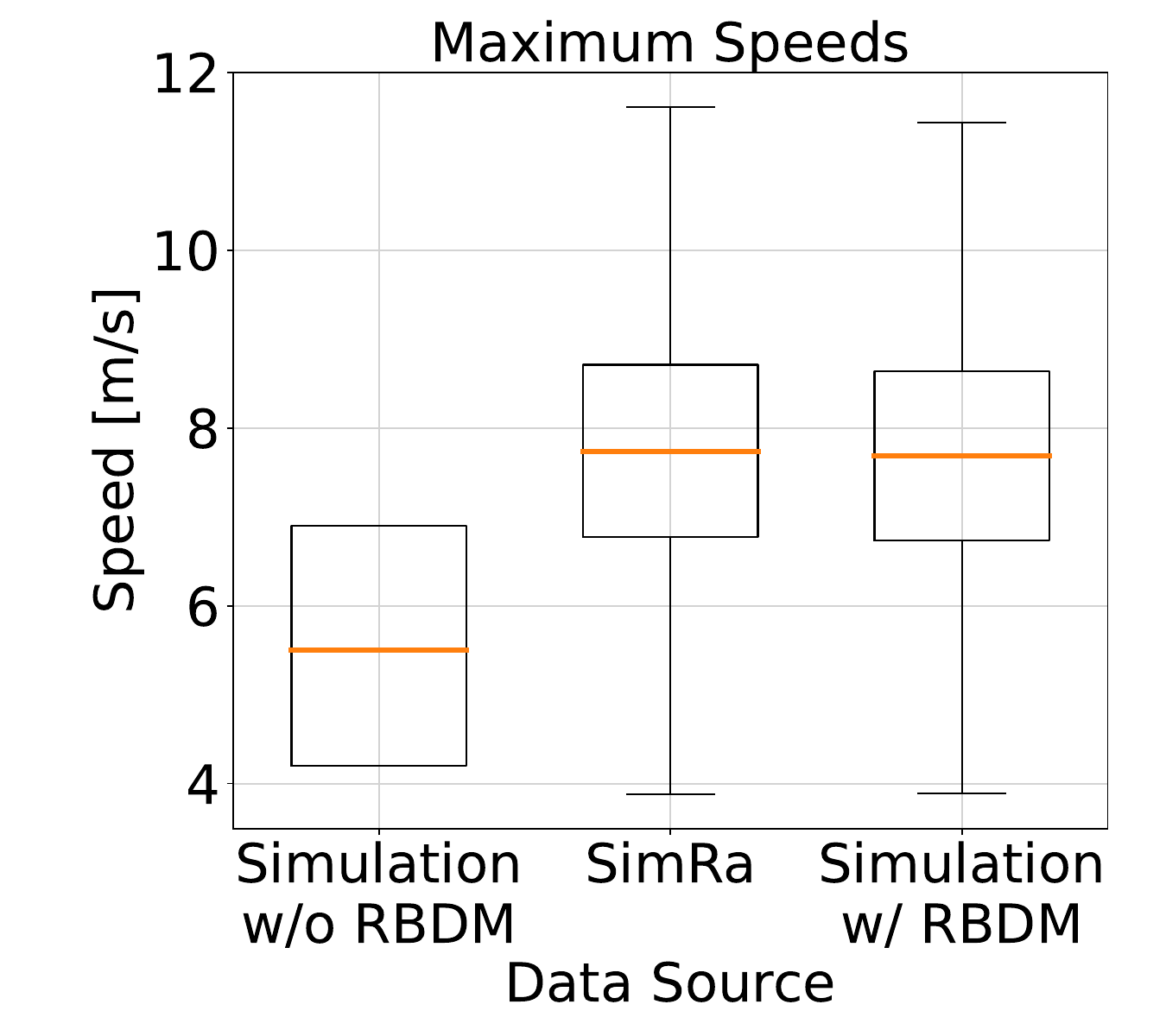}
    \caption{Maximum reached speed values of bicycles.}
    \label{fig:most-a}
\end{subfigure}
\hfill
\begin{subfigure}[t]{0.23\textwidth}
    \centering
    \includegraphics[width=\linewidth]{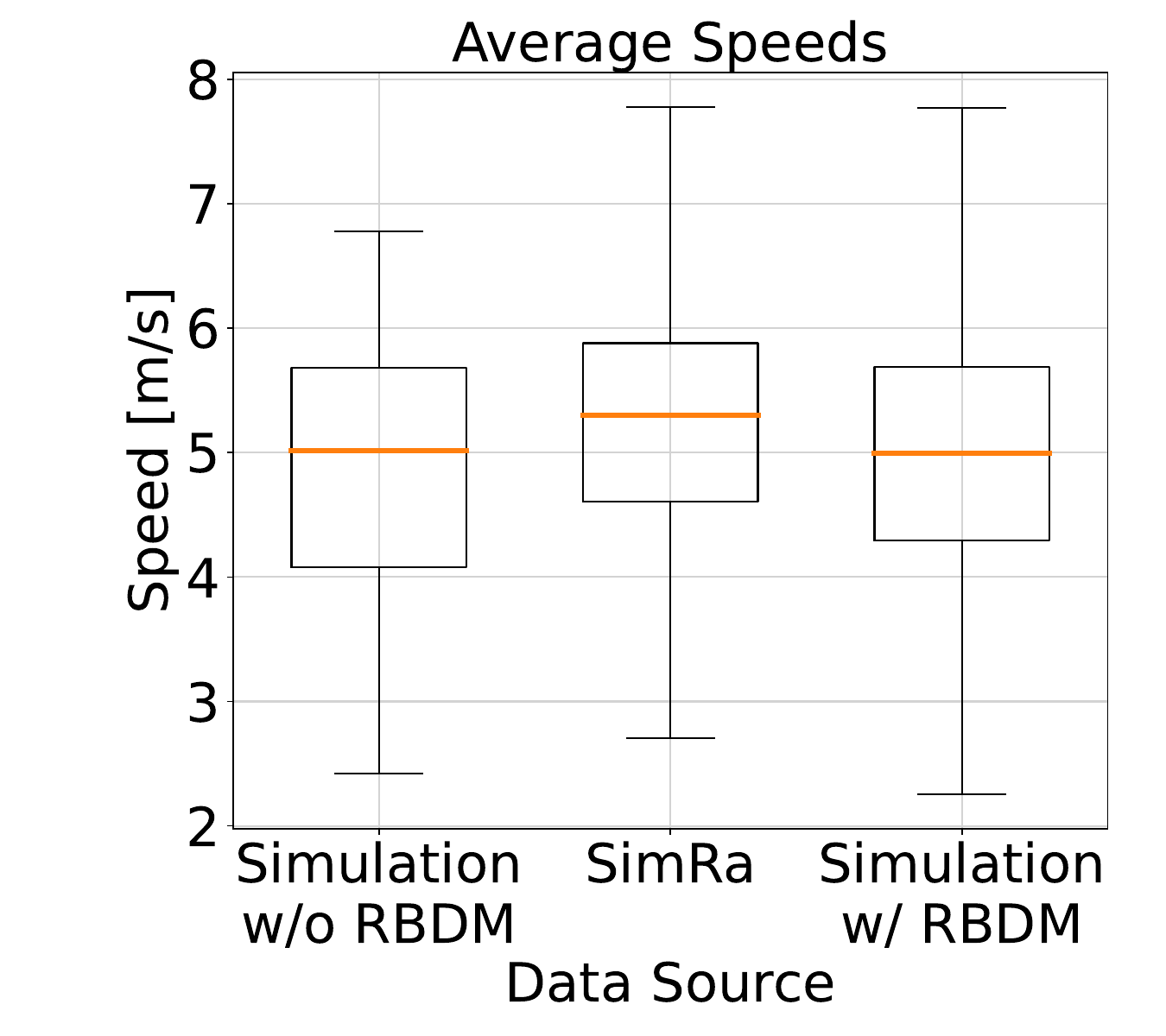}
    \caption{Average speed values of bicycles.}
    \label{fig:most-b}
\end{subfigure}
\hfill
\begin{subfigure}[t]{0.23\textwidth}
    \centering
    \includegraphics[width=\linewidth]{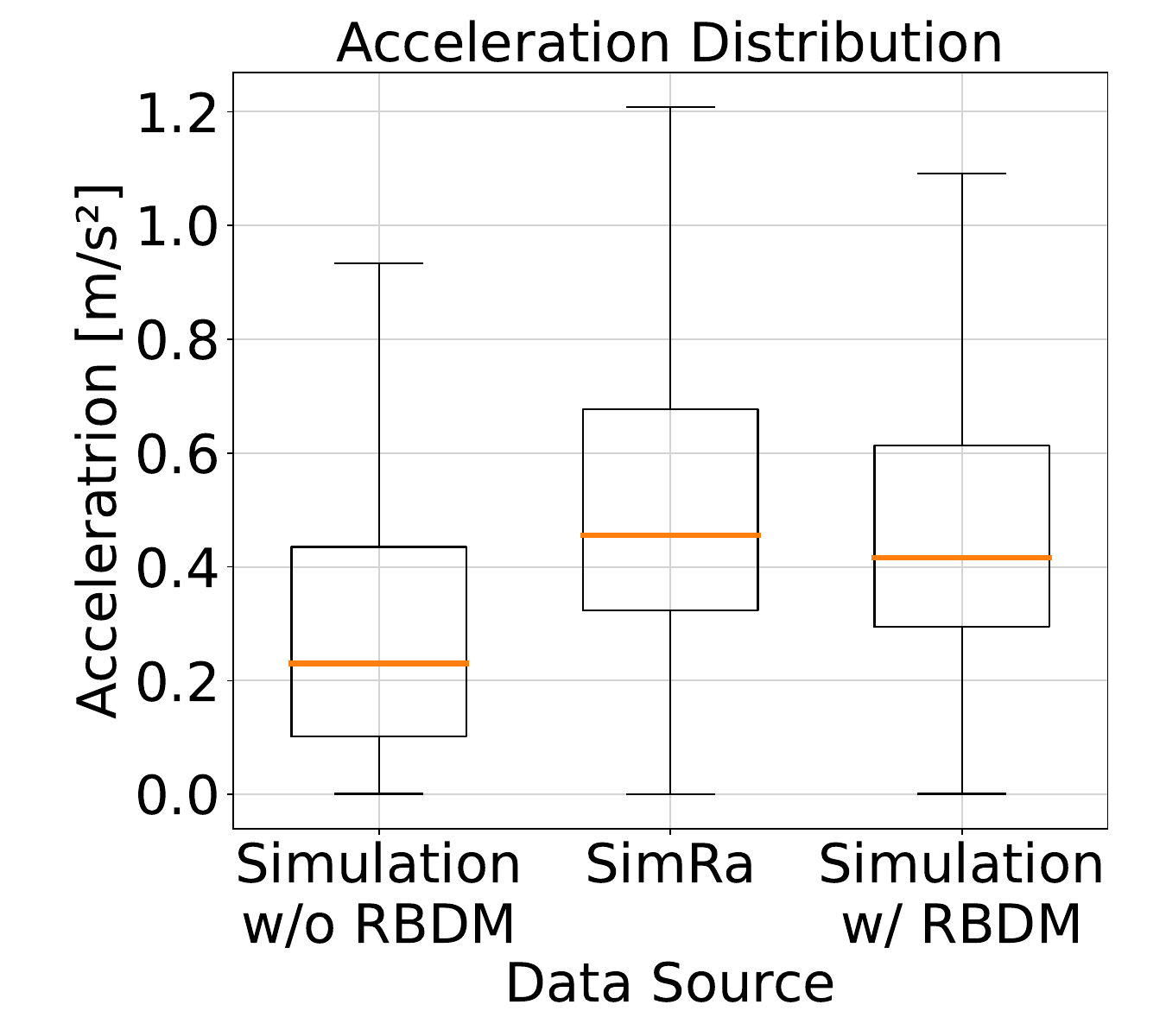}
    \caption{Distribution of acceleration of bicycles.}
    \label{fig:most-c}
\end{subfigure}
\hfill
\begin{subfigure}[t]{0.23\textwidth}
    \centering
    \includegraphics[width=\linewidth]{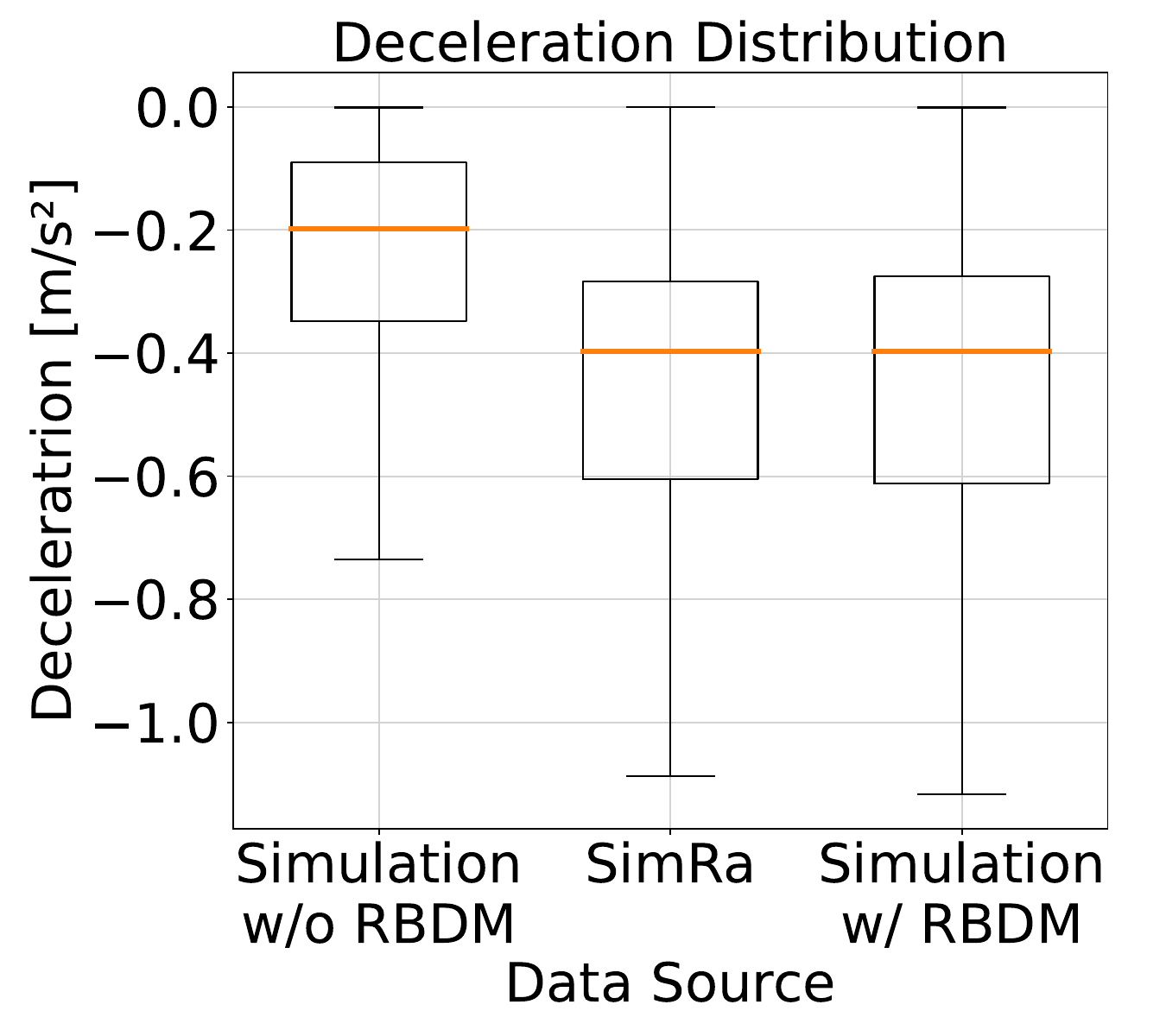}
    \caption{Distribution of deceleration of bicycles.}
    \label{fig:most-d}
\end{subfigure}
\caption{Results of MoST scenario evaluation with and without our model compared to the SimRa data results.}
\label{fig:most}
\end{figure}

In the subsequent evaluation, we assessed our model in the well-studied MoST scenario, with the results depicted in Figure  \ref{fig:most}. 
Similar to the Berlin scenario, we compared the MoST simulation with and without our RBDM against the SimRa data. 
Given the differing city topologies of Berlin and Monaco, we anticipated that the results might vary.

In Figure \ref{fig:most-a}, the maximum speeds for the different data sources are presented. 
The results closely resemble those of the Berlin scenario.
The RBDM demonstrated a closer alignment with the SimRa data, with a median maximum speed of $7.7\,m/s$, representing less than 1\% difference from the SimRa data ($7.73\,m/s$). 
In contrast, the simulation without our model showed a 29\% difference, with a median maximum speed of $5.5\,m/s$.
Also the 25th and 75th percentile results with the model closely align with the SimRa data, whereas the simulation without RBDM exhibits significant differences.

Figure \ref{fig:most-b} represents the average speeds of the different data sources, with the median values being very similar to the Berlin scenario.
$5.3\,m/s$ for SimRa, $4.99\,m/s$ for the simulation without RBDM, and $5.01\,m/s$ for the simulation with RBDM.
Also the 25th and 75th percentile and the IQR is very similar between the results with and without RBDM
However, the maximum average speed showed a notable difference, with our model reaching a maximum value of $9.12\,m/s$, closer to the SimRa data's $9.85\,m/s$, while the simulation without our model only reached $6.78\,m/s$.

Figure \ref{fig:most-c} and \ref{fig:most-d} represents the acceleration and deceleration distribution of the different data sources. 
The results are the same as for the Berlin scenario. 
Showing that the simulation with RBDM closely mirrored the SimRa data, while the simulation without RBDM was not able to produce realistic bicycle behavior.

\subsection{Discussion}
Overall, our RBDM demonstrated a closer alignment with the real-world data of the SimRa dataset compared to the simulations with the default model of SUMO, even in the well-known MoST scenario. 
The slight differences in speed distribution between the Berlin and MoST scenarios can be attributed to the distinct city topologies, with Berlin having more long and straight streets allowing for higher speeds compared to the shorter streets of Monaco. 
The evaluation results in MoST show that even in simulations of other cities, the RBDM will behave very similarly to the SimRa data set, which was recorded in Berlin. 
It is important to keep in mind that while using this model, it only represents bicyclists in Berlin and comparable cities.
Due to the absence of bicycle data from Monaco, we could not validate if bicyclists in Monaco will behave differently than those in Berlin. 
However, even in cities with different topologies than Berlin, we expect that our RBDM will lead to more realistic results than using the current model that represents bicycles as slow cars or fast pedestrians. 
A key fact is that our model is also easily extendable for other cities if data of those cities is available and if the bicycles will behave differently. 
For this it is only necessary to evaluate and change the PDF for the speed, deceleration, and acceleration. 

\section{Conclusion}
\label{sec::Conc}
\balance
This paper aimed to enhance the realism of bicycle behavior in the SUMO simulator by developing a new CF model using real-world data from the SimRa dataset. The focus was on improving the speed, acceleration, and deceleration behavior of bicycles. However, it is important to note that there are other factors that contribute to realistic bicycle movement, such as turning maneuvers, which were not fully addressed in this paper. Some of these dynamics are already developed by Karakaya et al. \cite{Karakaya2023Achieving}, as mentioned in the related work. Future work will involve integrating the models for these dynamics with our model.

Furthermore, future investigations could explore the inclusion of various types of cyclists. 
It is conceivable that different bicycle types exhibit distinct behaviors; for instance, racing bikes may demonstrate greater speed compared to city bikes, while eBikes may exhibit higher acceleration capabilities. 
To accommodate such distinctions, a larger volume of data pertaining to each cyclist group would be essential.

The developed CF model demonstrated superior performance in various scenarios compared to the current default model in SUMO. It notably improved acceleration, deceleration, and speed behavior. In particular, the maximum reached speed, acceleration and deceleration differed significantly between the two simulation setups, with the simulation including the RBDM closely resembling the SimRa data. In comparison, the average speeds of the bicycles were found to be very similar between the simulations with and without RBDM, depending on the scenario.

In conclusion, the introduced RBDM has the potential to significantly improve the accuracy of bicycle simulations in SUMO, particularly in the context of V2X communication, because the rules for generating messages mainly depend on these values. Additionally, it can also be beneficial in other areas such as infrastructure planning. Further research and development are needed to incorporate additional factors that affect bicycle movement and to continue refining the model to achieve even greater realism.

\bibliographystyle{IEEEtran}
\bibliography{bibliography}
\end{document}